\documentstyle[twocolumn,aps,epsf]{revtex}
\newcommand{\ignore}[1]{}

\begin{document}
\twocolumn[
\hsize\textwidth\columnwidth\hsize\csname @twocolumnfalse\endcsname
\draft
\title{ Surface EM waves in 1D Photonic Crystals} 
\author{J. Martorell}
\address {Dept.
  d'Estructura i Constituents de la Materia, Facultat F\'{\i}sica,\\
   University of Barcelona, Barcelona 08028, Spain
}
\author{D. W. L. Sprung and G. V. Morozov }
\address{
  Department of Physics and Astronomy, McMaster University\\
  Hamilton, Ontario L8S 4M1 Canada
}
\date{\today}
\maketitle
\begin{abstract}
Accurate analytic approximations are developed for the band gap 
boundaries and surface waves of a 1D photonic crystal, making use of 
the semiclassical theory recently developed by the authors. 
These analytic results provide useful insight on systematics of surface 
states. 
 \end{abstract}
\narrowtext
]

\section{Introduction}
Aside from their intrinsic interest, surface electromagnetic waves 
(SEW) have recently been proposed \cite{MMG04,MMG04b,Kra04,Zoz05} as 
a way to efficiently inject light into a photonic crystal waveguide, 
or to extract a focussed beam from a channel. Absent some such 
mechanism, the insertion and extraction is problematic. In the cited 
works, the photonic crystal (PC) was a two dimensional array of rods, 
of infinite length normal to the plane. In this paper we consider SEW 
on a one-dimensional (1D) PC, for which we recently developed accurate 
semi-classical approximations. 

Surface electromagnetic waves on a 1D PC were observed almost 30 
years ago \cite{Y2} \cite{Y3}. The basic theory was developed at that 
time, \cite{Y1P}, and can be found in the monograph of Yariv and Yeh 
\cite{Y1}. More recently the effect of varying the thickness of the 
termination layer has been measured experimentally \cite{Rob99,RM99} 
and a sensor based on the properties of SEW's has been proposed and 
demonstrated \cite{SR05}. In parallel, numerical calculations for 
SEW's in the bandgaps have been performed, and further aspects of the 
theory have also been developed \cite{RMH97,VGR03,VGA03,GAV04}. 

Here we study the properties of semi-infinite 1D photonic crystals 
with termination layers of various thicknesses. The dispersion 
equation for the SEW's is well known \cite{Y1}. However, exact 
results can be obtained only numerically, and to our knowledge, no 
simple analytic approximations for them have been developed.  
Accurate approximations not only assist in finding exact solutions, 
but also clarify the role of the various parameters defining the 
crystal. 

Starting from the dispersion equation,  we first derive exact 
expressions for the critical thicknesses at which SEW solutions 
appear and disappear. We then introduce approximate analytic forms 
for the dispersion relation which are valid in specified cases. We 
also apply the semiclassical method introduced by us in \cite{G1} and 
\cite{G2} to SEW's. These semiclassical approximations provide simple 
and accurate expressions for the bandgap edges. They also suggest a 
useful empirical parametrization that underlies our analytical 
approximate forms. 

For brevity we will discuss only the case of TE waves. Because our 
methods are formally quite different from those recently presented 
in \cite{GAV04}, in Section II we provide a short summary of the 
transfer matrix approach, in the notation of  our previous work 
\cite{G1}. In Section III we rederive the exact equation for SEW and 
obtain from it various analytic approximations for a semi-infinite 
crystal. With these in hand, we discuss systematics of SEW's. In 
Section IV we apply the semiclassical approximations of \cite{G1} to 
surface waves, and show that the second approximation is very 
accurate both for the dispersion relation and the bandgap boundaries. 
This lends support to the parametrization introduced in Section III. 
In  Appendix A we derive some closed analytic expressions for 
quantities introduced in \cite{G1} as infinite series. These allow a 
better understanding of the role of the various parameters of the PC.

 \section{Transfer matrix approach for a periodic crystal} 
We study surface waves of the Tamm type, that form at the interface 
between a uniform medium of low refractive index, $n_0$, and a 
semi-infinite 1-D photonic crystal, capped by a termination layer of  
the same material but selected width. To clearly separate the 
periodic bulk from the remaining surface layer, we split the 
termination layer of index of refraction $n_1$ and width $d_c$ into 
two pieces, of lengths $d_s + d_t = d_c$. Then the periodic array 
that makes the bulk of the crystal consists of ``cells'' each made of 
three uniform layers of widths $d_t$, $d_2$ and $d_1-d_t$ whose 
respective refraction indices are $n_1$ , $n_2$ and $n_1$. 
The cells are reflection symmetric when $2d_t = d_1$. A sketch 
is shown in Fig. \ref{fig01}. The initial cell, extending from $z=0$ 
to $z=d \equiv d_1+d_2 $ will be given the index $0$, the second 
index $1$, and so on, so that the p-th cell extends from $p d $ to 
$(p+1)d$ and has $n(z) = n_1$ when $p d <z < pd+d_t$ or $p d + 
d_t+d_2 < z < (p+1) d$ and $n(z)  = n_2$ when $p d+d_t <z < pd + 
d_t+d_2$.  As is customary, we will suppose $n_1 > n_2 > n_0$. The 
rest of the cap layer extends from $z=-d_s$ to $0$, and the uniform 
medium is located to the left of $z=-d_s$. 

We choose a coordinate system in which the layers have normal vector 
along OZ. An obliquely incident plane wave defines the OX axis. 
For monochromatic TE waves the electric field is parallel to the OY 
axis. As in \cite{G1}, we write 
 \begin{eqnarray}
{\bf E} &=& E(z) {\bf {\hat e}_y} e^{\displaystyle{i(k \beta x-\omega t)}} 
\nonumber \\
{\bf H} &=& \left[H_x(z) {\bf {\hat e}_x} +H_z(z){\bf {\hat e}_z}\right] 
e^{\displaystyle{i(k \beta x-\omega t)}}~,  
\label{eq:st1}
\end{eqnarray}
where $\omega$ is the angular frequency, $k=\omega/c$ is the vacuum 
wavenumber and $\beta k$ is the (constant) $x$-component of the wavevector of
 modulus $k(z)= 
\omega n(z)/c$. For a TE wave entering the  1-D photonic crystal from 
a uniform medium, one has 
 \begin{equation}
\beta = n_0 \sin \theta_0~, 
\label{eq:st2}
\end{equation}
where $\theta_0$ is the angle of incidence measured from the normal. 
Maxwell's equations require that in a periodic medium the $E(z)$ 
introduced in eq. \ref{eq:st1} satisfies 
 \begin{equation}
{{d^2E(z)}\over{dz^2}} + k^2( n^2(z) -\beta^2) E(z) = 0~, 
\label{eq:st3}
\end{equation}
The solutions are well known. In the $p$-th cell, the electric field 
corresponding to TE waves can be written as 
 \begin{equation}
E(z) = a_p e^{ik_1(z-pd)} + b_p e^{-ik_1(z-pd)} 
\label{eq:st4}
\end{equation}
when $ p d< z < p d+d_t$, and $k_1 = k \sqrt{n_1^2-\beta^2} \equiv k 
n_{1\beta}$. Similarly  
 \begin{equation}
E(z) = c_p e^{ik_2(z-pd)}+ d_p e^{-ik_2(z-pd)} 
\label{eq:st5}
\end{equation}
when $ pd+d_t < z < pd+ d_t+ d_2$, and $k_2 = k \sqrt{n_2^2-
\beta^2} \equiv  k n_{2\beta}$. Also, 
 \begin{equation}
E(z) = e_p e^{ik_1(z-pd)} + f_p e^{-ik_1(z-pd)} 
\label{eq:st6}
\end{equation}
when $ p d + d_t + d_2 < z < (p+ 1)d$. Matching these solutions and 
their derivatives at the two interfaces, one finds the transfer 
matrix, ${\cal M}$ [{\bf beware}: this matrix is called ${\cal M}^{-1}$ 
by some authors]. 
 \begin{equation}
\pmatrix{a_{p+1} \cr b_{p+1} } = {\cal M} \pmatrix{a_p \cr b_p}  
\equiv \pmatrix{A & B \cr B^* & A^*} \ \pmatrix{a_p \cr b_p}~,  
 \label{eq:st7}
\end{equation}
with
 \begin{eqnarray}
A &=& e^{ik_1 d_1} \left( \cos k_2 d_2 + {i\over 2} \left({k_1\over 
k_2} + {k_2 \over k_1} \right) \sin k_2 d_2 \right) \nonumber \\ 
 B &=& e^{i k_1(d_1 -2 d_t) } {i\over 2} \left({k_2 \over k_1}-
{k_1\over k_2}\right) \sin k_2 d_2~.  
 \label{eq:st8}
\end{eqnarray}
The periodic (Bloch) waves of the infinite crystal are eigensolutions 
of the transfer matrix 
 \begin{equation}
\pmatrix{a_{p+1} \cr b_{p+1}} = e^{\pm i\phi }\pmatrix{a_p \cr b_p} = 
{\cal M}  \pmatrix{a_p \cr b_p}~,  
 \label{eq:st9}
\end{equation} 
and therefore the secular equation is 
 \begin{equation}
(A - e^{\pm i \phi}) ( A^*-e^{\pm i\phi}) - |B|^2 = 0~, 
\label{eq:st10}
\end{equation}
with eigenvalues 
 \begin{equation}
e^{\pm i \phi} = {\rm Re}(A) \pm \sqrt{{\rm Re}(A)^2 - 1}~ ,
\label{eq:st11}
\end{equation}
where we have made use of det $ {\cal M} =|A|^2- |B|^2 = 1$. 
The corresponding eigenvectors are, up to a 
normalization factor 
 \begin{equation}
\pmatrix{ a \cr b} = \pmatrix{ B \cr e^{\pm i\phi} -A}~. 
\label{eq:st12}
\end{equation}
In allowed bands $\phi$ is real and the bandgap boundaries are 
defined by the condition Re$A = \pm 1$. In bandgaps the eigenvalues 
$\lambda_{\pm} \equiv e^{\pm i \phi}$ are real since 
$|{\rm Re} A| > 1$, and therefore 
 \begin{equation}
\lambda_{\pm} = {\rm Re}(A) \pm \sqrt{{\rm Re}(A)^2 - 1}~, 
\label{eq:st13}
\end{equation}
with $\lambda_- \lambda_+ = 1$. We choose the solution that gives a 
damped wave when $z \to + \infty$. It is $\lambda_-$ ( $\lambda_+$ ) 
when Re$(A) > 1$ ( $< -1$.) We write it as simply $\lambda$ and bear 
in mind that $|\lambda| < 1$. 

\section{Surface waves} 
These are states which decay in both directions, as one moves away 
from the surface of the photonic crystal. 
To the left of $z=0$ we have the variable portion of the cap layer, and 
the uniform external medium. The electric field is written as 
 \begin{equation}
E(z) = a_s e^{ik_1 z} + b_s e^{-ik_1 z}~, 
\label{eq:st14}
\end{equation}
when $-d_s < z <0$ and 
 \begin{equation}
E(z) = b_v e^{q_0 z}~, 
\label{eq:st15}
\end{equation}
with $q_0 = k \sqrt{\beta^2 - n_0^2}$, when $z < -d_s$. Matching 
at the interfaces, and choosing the damped wave solution for 
$z>0$, one finds 
 \begin{equation}
{q_0\over k_1} = -i {{\lambda -A - {\tilde B}}\over{\lambda -A + {\tilde B}}} 
 \label{eq:st16}
\end{equation}
where for simplicity we absorb a phase into 
 \begin{equation}
{\tilde B} \equiv e^{-2ik_1 d_s} B  \ .
 \label{eq:st17}
\end{equation}
Eq. \ref{eq:st16} determines the dispersion relation $k=k(\beta)$ for 
the surface waves.  It has to be solved numerically, and we will 
refer to the solutions thereby obtained as ``exact''. 

We begin by examining the structure of eq. \ref{eq:st16}. From the 
definitions of $q_0$ and $k_1$ given above, we see that the left hand 
side depends on $\beta$, but not on $k$. Eq. \ref{eq:st10} shows that 
$|\lambda -A| = |B| = |{\tilde B}|$, so that by writing 
 \begin{equation}
\theta_{\lambda-A} \equiv {\rm arg}(\lambda -A) \quad , \quad 
\theta_{\tilde B} = {\rm arg} ({\tilde B})~,  
 \label{eq:st18}
\end{equation}
eq. \ref{eq:st16} becomes:
 \begin{equation}
{q_0\over k_1} = -i {{e^{i \theta_{\lambda-A} } - e^{i \theta_{\tilde B}}}  
\over {e^{i \theta_{\lambda-A} } + e^{i \theta_{\tilde B}}}  } 
= \tan \left({{\theta_{\lambda-A} - \theta_{\tilde B}}\over 
2}\right)~. 
 \label{eq:st19}
\end{equation}
Next we look at arg$({\tilde B})$. From eq. \ref{eq:st8} we note that 
 \begin{equation}
{k_2 \over k_1}-{k_1\over k_2} = {n_{2\beta}\over n_{1\beta}}  - 
{n_{1\beta}\over n_{2\beta}} < 0~,  
 \label{eq:st20}
\end{equation}
according to the choice $n_1 > n_2$ made earlier. Hence, 
 \begin{equation}
\theta_{\tilde B} = {\pi \over 2} + k n_{1\beta} (d_1 -2 d_c) + \phi_s
\label{eq:st21}
\end{equation}
with $\phi_s$ chosen to be $0$ or $-\pi$ depending on the sign of 
$\sin k_2 d_2$. As shown in \cite{GAV04}, this sign is 
characteristic of each bandgap, unless the latter shrinks to zero 
width in what is called an optical hole \cite{GAV04}. Defining 
 \begin{equation}
\Theta(\beta) \equiv \tan^{-1} \left( \sqrt{ {\beta^2 -
n_0^2}\over{n_1^2 -\beta^2}}\right)~,
 \label{eq:st22}
\end{equation}
we can rewrite eq. \ref{eq:st19} as 
 \begin{equation}
\theta_{\lambda-A}(k) =  2  \Theta(\beta)+ k n_{1\beta} (d_1 -2 d_c) 
+ \phi_s + {\pi\over 2} + 2 \pi \nu 
 \label{eq:st23}
\end{equation} 
with $  \nu = 0, \pm 1, \pm 2, ... .$ . 
The l.h.s. and the second term on the r.h.s. of this equation 
depend on $k$, while the others do not. 

The term $2 \pi \nu$ arises from the inverse of the tangent appearing 
in eq. \ref{eq:st19}. When for a given $\beta$ and $d_c= d_{c,0}$  
the $\nu=0$ solution is $k=k_0$, one sees easily that $k=k_0$ is also 
a solution corresponding to the same $\beta$ and  $d_c = d_{c,0} + 
\nu \pi /(n_{1\beta_0} k_0)$, with $\nu = \pm 1, \pm 2, \cdots $. 
This is analogous to the well known property of solutions of the 
Schr\"odinger equation for a finite square well potential: increasing 
the width of the well by a half-wavelength, produces a state of the 
same energy but one additional node. For simplicity we will focus, 
from here on, on the case $\nu=0$. 

If $\theta_{\lambda- A}$ was linear or quadratic in $k$ for fixed 
$\beta$, we could easily solve eq. \ref{eq:st23} for $k(\beta, 
\cdots)$ in terms of the other parameters, and so identify the SEW's. 
Since that is not the case, one is reduced to numerical or graphical 
methods of solution. To see how this works, we consider an example in 
the first bandgap, taken from \cite{G1}: a PC with 
parameters $n_1= 2$, $d_1=100$ nm, $n_2= 1.5$ and $d_2= 250$ nm. In 
Fig. \ref{fig02} we plot separately the left hand side (continuous 
line), and the right hand side for several values of $d_c$ (dashed 
lines). In the first bandgap, $\sin k_2 d_2 > 0$ and therefore we have 
set $\phi_s = - \pi$.  As $k$ varies over the bandgap from $k_L$ to 
$k_R$, the argument $\theta_{\lambda-A}$ increases from $-\pi/2$ to 
$\pi/2$. This is a generic feature, as discussed in \cite{SMM04}. The 
intersection of any dashed line with the continuous line defines a 
corresponding solution for $k$. One sees that as $d_c$ decreases the 
corresponding $k$ increases, as expected (think of the analogy with 
the solutions of the Schr\"odinger equation). The extreme values of 
$d_c$ for which a solution exists will therefore be those for which 
the r.h.s. of eq. \ref{eq:st23} takes the values $\pm \pi/2$. This 
gives the values 
 \begin{eqnarray}
d_{c,min} &=& {d_1\over 2} + {1\over {k_R n_{1,\beta}}} \left( \Theta(\beta) - 
{\pi \over 2} \right) \nonumber \\ 
 d_{c,max} &=& {d_1\over 2} + {1\over {k_L n_{1,\beta}}}  \Theta(\beta)~. 
\label{eq:st24}
\end{eqnarray}
In the example shown in the figure, corresponding to $\beta = 1.2$, 
one finds $d_{c,min} = -27. $ nm and $d_{c,max} = 86.5 $ nm. The 
negative sign merely indicates that there is a surface wave solution 
for $d_c$ ranging from $0$ to $d_{c,max}$. 

To proceed further in the analysis of the SEW solutions requires 
values for the band edges $k_L,\, k_R$. These will be obtained from 
our semiclassical approximation \cite{G1} in Section IV. Before we 
delve into that, we first introduce an empirical approximation to 
$\theta_{\lambda-A}$ which will be justified by the semiclassical 
theory. For the first bandgap, we write 
 \begin{equation}
\theta_{\lambda-A}^{(e)} = \sin^{-1}\left({{k-k_m}\over {w/2}}\right) 
\label{eq:st25}
\end{equation}
 with
$ k_m \equiv (k_R + k_L)/2$ and $w = k_R-k_L$.  
In Fig. \ref{fig03} we compare the exact and the empirical forms of 
$\theta_{\lambda-A}$ for four values of 
$\beta$ ranging from $n_0$ to $n_2$. It can be seen that the 
approximation is quite satisfactory. 

Based on the above development, we now derive two analytical 
approximations for the dispersion relation. The first is valid when 
the crossing point in Fig. \ref{fig02} lies in the linear portion of 
$\theta_{\lambda-A}$, while the second is valid near the upper and 
lower ends of the curve. 


\subsection{Solutions in the middle of the bandgap.} 

These are of particular interest because the damping is stronger, 
strongly confining the wave to the surface \cite{Rob99}.  

Assuming that $k-k_m << w$ we can approximate $\sin^{-1} (2(k-k_m)/w) 
\simeq 2(k-k_m)/w$ and eq. \ref{eq:st23} then has a solution 
 \begin{equation}
k \simeq {{ \Theta(\beta) +   k_m/w + \phi_s/2 + \pi (\nu +1/4)}\over 
{1/w + n_{1\beta}( d_c -d_1/2 )}}  \ . 
 \label{eq:st26}
\end{equation}
In this way we can easily construct $k( \beta)$ for a fixed 
value of $d_c$, or conversely, study $k = k(d_c)$ for fixed $\beta$. 
The role of the bandgap parameters $k_m$ and $w$ is also quite easy 
to analyze. Fig. \ref{fig04} shows the accuracy of this 
approximation when $d_c = 25$ nm. For this example, when $\beta > 1.4 
$ the approximation ceases to be valid and one has to resort to other 
approximations described in the next subsection. 

\subsection { Solutions near the bandgap boundaries.}

These approximations will be useful in analyzing the results of 
experiments like those of Robertson and May \cite{RM99}, where the 
SEW's appear very close to the boundaries. As seen in Fig. 
\ref{fig03}, the linear approximation to the arcsine fails near 
the band boundaries. We discuss solutions near the lower boundary, 
but similar approximations can be developed for the upper boundary. 

When $k$ is slightly above $k_L = k_m-w/2$, it is convenient to 
introduce $\zeta > 0$ via  
 \begin{equation}
k- k_m = -{w\over 2}(1-\zeta) \ . 
\label{eq:st27}
\end{equation}
Then 
 \begin{eqnarray}
\sin^{-1} \left( {{k-k_m}\over {w/2}} \right) &\simeq& - {\pi \over 2} + 
\sqrt{2\zeta} \nonumber \\ 
&=& -{\pi \over 2} + 2 \ \sqrt{{k-k_L}\over w}~. 
 \label{eq:st28}
\end{eqnarray} 
Inserting this into eq. \ref{eq:st23} gives 
 \begin{equation}
2 \ \sqrt{{k-k_L}\over w} + p (k-k_L) = \Lambda 
\label{eq:st29}
\end{equation}
with
 \begin{eqnarray}
p &\equiv& n_{1\beta} ( 2 d_c -d_1) \nonumber \\
 \Lambda &\equiv& 2 \Theta + 
\phi_s + \pi ( 2 \nu +1) - k_L p \ , 
 \label{eq:st30}
\end{eqnarray}
and solving for $k - k_L$ 
 \begin{equation}
k = k_L + {1\over { p^2}}\left( -{1\over \sqrt{ w}} + \sqrt{{1\over w} 
+  p \Lambda} \right)^2 \ . 
 \label{eq:st31}
\end{equation}
which is the desired solution, $k = k(\beta)$, near the lower bandgap 
boundary. Fig. \ref{fig05} shows an example of the accuracy of this 
expression. Furthermore, when $p \Lambda $ is small 
compared to $1/w$ one can expand and find 
 \begin{equation}
k = k_L + {1\over 4} w \Lambda ^2  \ ,
\label{eq:st32}
\end{equation}
which again manifests the dependence of $k-k_L$ on $d_c$ and $w$, 
and allows one to construct $k=k(\beta)$ very easily. Fig. \ref{fig05}  
shows again the validity of this approximation. Note also that the 
condition $\Lambda=0$ determines the location of the zone boundary. 
Writing it out, one recovers eq. \ref{eq:st24}, so that eqs. 
\ref{eq:st31} and \ref{eq:st32} do not violate this exact relation.

\subsection{ Surface states in the second bandgap}

In the examples discussed above we focussed on dispersion relations 
for SEW in the first bandgap. The specific system considered has no 
optical holes. However, it does have an optical hole in the second 
bandgap, so we examine this situation to clarify  what that entails. 

In eq. \ref{eq:st23}, the angle $\theta_{\lambda-A} $ ranges from $-\pi/2 
+ \pi (q -1)$ to $\pi/2 + \pi(q-1)$ in the $q$-th bandgap.  Following 
the previous argument we find the critical thicknesses for 
appearance/disappearance of SEW's to be 
 \begin{eqnarray}
d_{c,min} &=& {d_1\over 2} + {1\over {k_R n_{1,\beta}}} 
\left(\Theta(\beta) + {\phi_s\over 2} - {\pi \over 2}(q-1) + \pi \nu\right) 
\nonumber \\
d_{c,max} &=& {d_1\over 2} + {1\over {k_L n_{1,\beta}}} \left(\Theta(\beta) + 
{\phi_s\over 2} - {\pi \over 2}(q-2) + \pi \nu\right) \nonumber \\
\ .
\label{eq:st33}
\end{eqnarray}
In the second bandgap, Fig. \ref{fig06} shows that an optical hole 
occurs, where the width of the gap shrinks to zero. (See Section IV  
and \cite{GAV04} for the exact location.) At the optical hole, $k_2 
d_2 = \pi$, giving a change of sign of $\sin k_2 d_2$ at this 
critical $\beta=\beta_{oh}$. Correspondingly, one finds that $\phi_s 
= 0$ for $\beta <\beta_{oh}$ and $ \phi_s = - \pi$ for $\beta> 
\beta_{oh}$. The continuous lines in Fig. \ref{fig07} show 
$d_{c,max}$ and $d_{c,min}$, computed from eq. \ref{eq:st33} with 
$\nu=0$.  The horizontal dashed lines correspond to various values of 
$d_c$. The plot shows that for $d_c = 10$ nm, there is a SEW for 
$\beta$ less than approximately $1.25$, and that the SEW reappears 
for $\beta > \beta_{oh}.$ Similarly, for $d_c = 15$ nm there is a SEW 
when $\beta$ is less than approximately 1.325, reappearing for $\beta 
> \beta_{oh}$. For $d_c = 20$ nm SEW appear for $\beta < \beta_{oh}$, 
and for $\beta$ greater than approximately 1.4. Finally, for $d_c = 
40$ nm there is a SEW only when $\beta < \beta_{oh}$.  

Fig. \ref{fig07} shows surface wave solutions when $d_c = 0.15 
d_1$ and $d_c = 0.40\ d_1$, and  confirms the above discussion. Note 
that numerically it is rather difficult to locate the end point of 
the surface wave solution for $d_c = 0.15\ d_1$, because the 
solutions run very close to the band boundary. In contrast, eq. 
\ref{eq:st33} and the graph Fig. \ref{fig07} locate that end point 
very easily. 

\subsection{ Solutions when $\beta > n_2$}
In this regime, $k_2 = k \sqrt{n_2^2 -\beta^2}$ becomes imaginary and 
we write $k_2 = i q_2$. The expressions for $A$ and 
${\tilde B}$ become  
 \begin{eqnarray}
A &=& e^{ik_1 d_1} \left( \cosh q_2 d_2 + {i\over 2} \left({k_1\over 
q_2}-{q_2\over k_1} \right) \sinh q_2 d_2 \right) \nonumber \\ 
 {\tilde B} &=& - {i\over 2} e^{ik_1 (d_1 - 2 d_c)}\left({k_1\over 
q_2}+{q_2\over k_1} \right)  \sinh q_2 d_2 \ . 
 \label{eq:st34}
\end{eqnarray}
As $\beta$ increases, the hyperbolic functions soon become 
very large, giving 
 \begin{eqnarray}
A &\simeq & {1\over 2} e^{i k_1 d_1 + q_2 d_2} \left( 1+ {i\over 2} 
\left({k_1\over q_2}-{q_2\over k_1} \right)\right) \nonumber \\ 
 {\tilde B} &\simeq& -{i\over 4} e^{ik_1(d_1-2d_c)+ q_2 d_2} 
\left({k_1\over q_2}+{q_2\over k_1} \right) \ ,
 \label{eq:st35}
\end{eqnarray}
and therefore:
 \begin{equation}
{\rm Re}(A) \simeq {1\over 2} e^{q_2 d_2} (\cos k_1 d_1-\Gamma \sin k_1 d_1)
\label{eq:st36}
\end{equation}
with
 \begin{eqnarray}
\Gamma &\equiv& {1\over 2} \left({k_1\over q_2}-{q_2\over k_1} \right)
= 
{1\over 2}\ {{n_1^2+ n_2^2 - 2\beta^2}\over{\sqrt{(n_1^2-
\beta^2)(\beta^2-n_2^2)}}} \  ,
 \label{eq:st37}
\end{eqnarray}
which is independent of $k$. The stationary points of the quantity in
parentheses on the right of eq. \ref{eq:st36}, are at $\tan k_1 d_1 = 
- \Gamma$. At these points: 
 \begin{equation}
{\rm Re}(A) = (-)^q {1\over 2} e^{q_2 d_2} \sqrt{1+ \Gamma^2} 
\label{eq:st38}
\end{equation}
Therefore Re$(A)$ alternates in sign from one bandgap to the 
next, and the amplitude of the oscillations is large in most of 
the range of values of $q_2$ due to the exponential factor. Writing 
Re$(A) = (-1)^q |{\rm Re}(A)|$, and using eq. \ref{eq:st13}, we 
obtain 
 \begin{eqnarray}
\lambda &=& (-)^q \left(|{\rm Re}(A)| - \sqrt{ ({\rm Re}(A))^2 - 
1}\right) 
\simeq {{(-)^q}\over {2 |{\rm Re}(A)|}} \ , 
 \label{eq:st39}
\end{eqnarray}
which is much smaller than $|A|$ over most of the bandgap. Therefore 
 \begin{equation}
\theta_{\lambda -A} \simeq {\rm arg} (-A) = - \pi + k_1 d_1 + 
\tan^{-1} \Gamma \ , 
 \label{eq:st40}
\end{equation}
where the last term is independent of $k$ and therefore  
$\theta_{\lambda-A} $ becomes linear in $k$. Figure  \ref{fig09} shows 
the accuracy of eq. \ref{eq:st40}: beyond $\beta = 1.6$, the 
exact (continuous) and the approximate (dashed) dispersion relations 
practically coincide. Using this approximation and equating 
$\theta_{\lambda -A}$ to $(q -1)\pi \pm \pi/2$ we get an expression for the 
bandgap boundaries  
 \begin{equation}
k_{L,R} = {1\over {n_{1\beta} d_1}} \left[ q\pi \pm{\pi \over 2} \nonumber \\
- \tan^{-1} \Gamma \right]
 \label{eq:st41}
 \end{equation}
Figure \ref{fig10} shows that the band boundaries are quite 
well reproduced by this approximation. The exception is the lowest boundary 
when $\beta < 1.6$. We can also use 
the approximation for $\theta_{\lambda-A}$ to predict the dispersion 
relation:  inserting \ref{eq:st40} into eq. \ref{eq:st23} and 
solving for $k$, one finds 
 \begin{equation}
k = {1\over{n_{1\beta} d_c}} \left[ \Theta + \nu \pi + 3 {\pi \over 4} 
+ {\phi_s\over 2} 
-{1\over 2} \tan^{-1} \Gamma \right] \ .
\label{eq:st42}
\end{equation}
As  figure \ref{fig10} shows, the exact and
the approximate curves for $k=k(\beta)$ are very close (indistinguishable
when $\beta > 1.6$). Again, eq. \ref{eq:st42} shows very explicitly the
role of $d_c$ and of the indices of refraction in determining $k$.

\section{Semiclassical approximations}

The elements of the transfer matrix can be related to the amplitudes 
for transmission $t_B$ and reflection $r_B$, for a single cell: 
 \begin{equation}
\pmatrix{t_B \cr 0} = \pmatrix{ A & B \cr B^* & A^*} \pmatrix{1 \cr r_B}   \ ,
\label{eq:st43}
\end{equation}
that is,  $A = 1/t^*_B$ and $B = - A r^*_B$ . The semiclassical
expressions for $r_B$ and $t_B$ are found in eqs. 35 and 36 of \cite{G1} for
the first and second approximations discussed in that reference.

\subsection{ First approximation} 
In this approximation, and for a single cell, eq. 35  of \cite{G1} gives 
 \begin{eqnarray}
r_B^{(1)} &=& {{-s_q^* \sinh (\gamma_1 d) }\over{\gamma_1 \cosh(\gamma_1 d) 
-i \delta_q \sinh(\gamma_1 d)}} \nonumber \\
t_B^{(1)} &=& {{(-)^q \gamma_1}\over{\gamma_1 \cosh (\gamma_1 d) - i \delta_q 
\sinh(\gamma_1 d)}} \ .
\label{eq:st44}
\end{eqnarray}
Therefore
 \begin{eqnarray}
A_s^{(1)} &=& (-)^q \left( \cosh \gamma_1 d+ i{\delta_q \over \gamma_1} 
\sinh \gamma_1 d\right) \nonumber \\
B_s^{(1)} &=& (-)^q {s_q \over \gamma_1 } \sinh \gamma_1 d \ ,
\label{eq:st45}
\end{eqnarray}
where the subscript $s$ stands for  for semiclassical, and the 
superscript $(1)$ denotes the first approximation introduced in 
\cite{G1}. The explicit expression for $s_q$ is given below, eq. 
\ref{eq:st46}, $\gamma_1 \equiv \sqrt{|s_q|^2- \delta_q^2}$ and the 
detuning from the q-th Bragg resonance is $\delta_q = n_{av,\beta} 
(k-k_B)$, with $k_B = q \pi/(n_{av,\beta} d)$ . In addition, 
$\gamma_1 d$ is the exponent of the damped wave constant, $\lambda$, 
as we can easily confirm: inserting Re$ A^{(1)}_s$ into eq. 
\ref{eq:st13}  we find $\lambda_s^{(1)} = (-)^q e^{-\gamma_1 d}$. 

 Using expressions from \cite{G1}, for an asymmetric cell we 
can write $s_q$ in compact form 
 \begin{eqnarray}
s_q &=& -{i\over d} \ln{n_{1\beta}\over n_{2\beta}} e^{i(\phi_1+\phi_2)/2}
\sin((\phi_1-\phi_2)/2) 
\label{eq:st46} \\ 
{\rm with} \quad 
\phi_1 &=& 2 ( -k_1 + \delta_q )d_t \nonumber \\
\phi_2 &=& 2 ( -k_1 d_t -k_2 d_2 + \delta_q (d_t+d_2)) \ .
\label{eq:st47}
\end{eqnarray}
From eq. \ref{eq:st45} we find 
 \begin{equation}
\lambda_s^{(1)}- A_s^{(1)} = (-)^{(q-1)} \sinh(\gamma_1 d) \left( 1 + 
i {\delta_q \over \gamma_1} \right)  \ ,
\label{eq:st48}
\end{equation}
and therefore $\theta_{\lambda_s^{(1)}-A_s^{(1)}}$ 
\begin{eqnarray}
& & \qquad \equiv {\rm arg}(\lambda_s^{(1)} 
-A_s^{(1)}) = (q-1)\pi +\tan^{-1} {\delta_q \over \gamma_1} \nonumber \\ 
&=& (q-1) \pi+ \sin^{-1} {\delta_q \over{|s_q|}} 
\simeq (q-1)\pi + \sin^{-1}  {{\delta_q} \over {S_B}}~, 
\label{eq:st49}
\end{eqnarray} 
where in the last step we have replaced the slowly varying $|s_q(k)|$ 
by $S_B$, its value at the Bragg resonance $k=k_B$. Since by 
definition, $\delta_q = (k-k_B) n_{av,\beta}$, we arrive at
 \begin{equation}
\theta_{\lambda_s^{(1)}-A_s^{(1)}} \simeq (q-1)\pi + \sin^{-1} \left({{k-k_B}
\over {w/2}}\right) \ ,
\label{eq:st50}
\end{equation}
with $w= 2 S_B/n_{av,\beta}$. This has the same form as the empirical 
parametrization used above, but provides explicit estimates for 
the width and position $k_m = k_B$ of the bandgap.  

\subsection{ Second approximation} 

The second, and more accurate, approximation introduced in \cite{G1} 
leads to similar but more involved expressions for $A^{(2)}$ and 
$B^{(2)}$. From eq. 36 of that reference, we find 
 \begin{eqnarray} 
A_s^{(2)} &=& (-)^q \bigg[ \cosh \gamma_2 d \nonumber \\ 
&+& i {{\left[(1+|u|^2)\eta_q - 2 {\rm Im}(s_q u^*)\right] }\over{(1-|u|^2
)\gamma_2}} \ \sinh \gamma_2 d \bigg] \nonumber \\ 
B_s^{(2)} &=& (-)^q \ {{s_q -2 i \eta_q u -s_q^* u^2}\over{(1-|u|^2)
\gamma_2 }} \sinh \gamma_2 d \ . 
\label{eq:st51} 
\end{eqnarray} 
where $\gamma_2 \equiv \sqrt{|s_q|^2- \eta_q^2}$ and $\eta_q = \delta_q - 
i c_2$. In the appendix we give new analytic expressions for 
$u=v_1(z=0)$ and for $c_2$, that sum the series written in \cite{G1}, 
eqs. 18 and 19. As in the previous case, one easily finds that 
$\lambda_s^{(2)} = (-)^q e^{-\gamma_2 d}$.\\ 

Inserting $A_s^{(2)}$ and $B_s^{(2)}$ into eq. \ref{eq:st16}, and 
solving, one finds the corresponding predictions for $k=k(\beta)$. 
Before presenting these results, we will show the usefulness of this 
second approximation in giving accurate values for the bandgap 
boundaries. 

In eqs. (31) and (32) of ref. \cite{G1} we showed that the first 
approximation provides simple estimates for the bandgap boundaries. 
In the second approximation, the condition $\gamma_2 =0$ defines the 
band boundaries, because it corresponds to infinite decay length of 
the surface state.  Using the explicit form of $\gamma_2$ from eq.  
(25) of \cite{G1}, we can write this condition, for the $q$-th 
bandgap boundaries, as 
 \begin{eqnarray}
-|s_q(k_L)| &=& (k_L-k_B) n_{av,\beta} + r_2 \nonumber \\
|s_q(k_R)| &=& (k_R-k_B) n_{av,\beta} + r_2  \ ,
\label{eq:st52}
\end{eqnarray}
where $r_2 \equiv -i c_2$. The dependence of $|s_q|$ on $k$ 
is fairly smooth as long as $k$ remains within a bandgap. To a good 
approximation one can expand around $k=k_B$ and write 
 \begin{equation}
|s_q(k)| = S_B + \xi (k-k_B) -{1\over 2} \eta (k-k_B)^2 \ ,
\label{eq:st53}
\end{equation}
with
 \begin{eqnarray}
S_B &\equiv& |s_q(k_B)| ={1\over d} \ln\left({n_{1,\beta}\over 
n_{2,\beta}}\right) \sin \alpha_B \ .\nonumber \\ 
 \alpha_B &\equiv& \pi q {{n_{2,\beta} d_2}\over{n_{av,\beta} d}} 
\nonumber \\ 
 \rho &\equiv& {{d_1d_2}\over d} (n_{2,\beta}-n_{1,\beta}) \nonumber \\ 
 \xi &\equiv& {1\over d} \ln\left({n_{1,\beta}\over 
n_{2,\beta}}\right) \rho \ \cos \alpha_B \nonumber \\ 
 \eta &\equiv& {1\over d} \ln\left({n_{1,\beta}\over 
n_{2,\beta}}\right) \rho^2 \sin \alpha_B \ . 
 \label{eq:st54}
\end{eqnarray}
Inserting this expansion into each line of \ref{eq:st52} leads to 
the desired analytic expressions 
 \begin{eqnarray}
k_L &=& k_B + {1\over \eta} \bigg[n_{av,\beta}+ \xi \nonumber \\ 
&&  \qquad \qquad - \,\,
\sqrt{(n_{av,\beta}+ \xi)^2 + 2 \eta (S_B+r_2)}\bigg] \nonumber \\ 
k_R &=&  k_B  + {1\over \eta} \bigg[-n_{av,\beta}+ \xi \nonumber \\ 
&&  \qquad \qquad + \,\,
\sqrt{(n_{av,\beta}- \xi)^2 + 2 \eta (S_B-r_2)}\bigg]  
\label{eq:st55}
\end{eqnarray}
with $r_2$ evaluated at $k=k_B$:
 \begin{equation}
r_2 = {{d_1-d_2}\over{4d^2}} \left(\ln\left({n_{1,\beta}\over 
n_{2,\beta}}\right) \right)^2 \sin( 2 \alpha_B) \ . 
\label{eq:st56}
\end{equation}
In Fig. \ref{fig06} we compare the exact bandgap boundaries to those 
of eq. \ref{eq:st55} and, as can be seen, the latter works very well 
except very close to $\beta = n_2$, where the semiclassical 
approximation is expected to fail. For $\beta$ close to $n_0$ the 
values of $\eta$ are small and one can neglect them in eq. 
\ref{eq:st53}. This leads to simpler forms: 
 \begin{eqnarray}
k_L &=& k_B - {{S_B+r_2}\over {n_{av,\beta}+ \xi}} \nonumber \\
k_R &=& k_B + {{S_B-r_2}\over {n_{av,\beta}- \xi}}  \ .
\label{eq:st57}
\end{eqnarray}
Neglecting the contribution from $r_2$ we recover the expressions 
of the first approximation ($\gamma_1 = 0$) already discussed in 
\cite{G1}. 

Direct inspection of the expression for Re$(A)$ derived from  eq. 
\ref{eq:st8} shows that the points where the bandgaps shrink to zero 
width correspond to special values of the optical depths of the 
layers: $k_1 d_1 = m_1\pi$, $ k_2 d_2 = m_2 \pi$, with $m_1, m_2$ 
integers. These give 
 \begin{equation}
n_{1,\beta} d_1 m_2 = n_{2,\beta} d_2 m_1 \ ,
\label{eq:st58}
\end{equation}
so that:
 \begin{equation}
\beta_{oh}^2 = {{n_1^2 d_1^2 m_2^2- n_2^2 d_2^2 m_1^2}\over{d_1^2 
m_2^2- d_2^2 m_1^2}}   
 \label{eq:st59}
\end{equation}
and $ k_{oh} = m_1 \pi/(d_1 n_{1,\beta_{oh}})$, in agreement with 
\cite{GAV04}. The point where the second bandgap shrinks to zero 
width, in Fig. \ref{fig06}, corresponds to $m_1=m_2=1$, $q=2$. It is 
easy to check that with the above semiclassical approximations one 
finds exactly the same optical hole. Note that at this critical value 
 \begin{equation}
\alpha_B = \pi q {{k_2 d_2}\over{k_1 d_1 + k_2 d_2}} = \pi q {{m_2 
\over{m_1+m_2}}} = \pi 
 \label{eq:st60}
\end{equation}
so that $r_2=0$, $S_B = \eta=0$, and therefore
 \begin{equation}
k_L = k_R = k_B \ .
\label{eq:st61}
\end{equation}

\subsection{ Results for SEW's}
  In Fig. \ref{fig11} we compare the exact and semiclassical results 
for $k=k(\beta)$, choosing three thicknesses for the cap layer: 
$d_c=d_t = 25, 50$ and $75$ nm. The first approximation becomes 
inaccurate when $\beta$ exceeds $1.4$, but gives accurate results up 
to that value. The second approximation is so close to the exact 
values that one can see the difference only for values of $\beta$ 
very close to the critical value $\beta = n_2 = 1.5$. Beyond that, 
our semiclassical approximations cannot be applied, since $k_2$ 
becomes imaginary. In the second bandgap, the accuracy of the first 
approximation is significantly worse, whereas the second 
approximation is as good as for the first bandgap. For brevity, we do 
not show any figures for $A$ and $B$ as functions of $k$ and $\beta$. 
For most of the values in the first bandgap the agreement is similar to 
that seen in fig. \ref{fig11} for the dispersion relations.

\section{Summary and Conclusions}
By considering the semi-infinite limit of a 1D PC we have derived a 
dispersion equation for SEW's, valid for termination layers of 
selected width. Our goal has been to clarify the systematics of the 
solutions of this equation. To do so we first discussed a 
graphical solution. This allowed us to derive analytic 
expressions for the critical thicknesses at which solutions 
appear at the bandgap boundaries. Further, by introducing a suitable 
parametrization of $\theta_{\lambda-A}$, \ref{eq:st25}, we have 
derived simple approximations for solutions either in the middle of 
the bandgap or near the edges: eqs. \ref{eq:st26}, \ref{eq:st31}, and 
\ref{eq:st32}. We tested them by an example whose first bandgap has 
no optical holes. We then extended the analysis of critical 
thicknesses to the second bandgap, where there {\it is} an optical hole. 
The appearance and disappearance of solutions as a function of cap 
layer thickness is again easily predicted: eq. \ref{eq:st33}. For 
completeness we have also examined solutions when $\beta > n_2$, and 
again found very simple approximations eqs. \ref{eq:st41} and 
\ref{eq:st42}, valid over most of the range $n_2 < \beta < n_1$, that 
allow us to study the systematics in a transparent way.

Finally, in section IV, we applied the semiclassical approximations 
derived in \cite{G1} to SEW's. The second approximation is very 
accurate in predicting the bandgap boundaries and the dispersion 
relation. We took advantage of this to derive accurate and simple 
approximations for the boundaries: eqs.  \ref{eq:st56} and 
\ref{eq:st57}. The first approximation already supports the validity 
of the empirical parametrization of $\theta_{\lambda-A}$: {\it cf.} 
eqs. \ref{eq:st25} and \ref{eq:st50}.  

 In conclusion: we have presented a set of analytic results, exact 
and approximate, that clarify the systematics of solutions for 
surface EM waves in semi-infinite 1D photonic crystals. In addition, 
we have found simple analytic expressions for the bandgap boundaries 
that should be useful for the design of PC configurations. We plan 
next to extend our results to layered configurations with cylindrical 
symmetry \cite{XLY00,Iba03}. It would also be interesting to see whether 
the systematics found here apply to surface states in 2D and/or 3D 
photonic crystals.

\section{Acknowledgements}

We are grateful to NSERC-Canada for Discovery Grant RGPIN-3198 
(DWLS), and to DGES-Spain for continued support through grants  
BFM2001-3710 and FIS2004-03156 (JM).

\appendix
\section{Closed forms for series used in the semiclassical approximations}

\subsection{Analytic expression for $u=v_1(0)$}
The function $v_1(z)$ is defined in eq. 19 of \cite{G1}  as 
 \begin{equation}
v_1(z) = -{{ id}\over {2\pi}} \ \sum_{m \ne q} {{s_m 
e^{{i 2\pi(m-q) z /d}}}\over{m-q- \delta_q d/\pi}}
\label{eq:aa01}
\end{equation}
with $s_m$ given in eq. 12 of \cite{G1}. In the case of interest here, 
 the refractive index is piece-wise {\it constant}, so that 
 \begin{eqnarray}
s_m &=& {1 \over {2d}} \sum_j 
\ln{{n_e(z_j+0)}\over{ n_e(z_j-0)}} \ e^{{2i(-
\psi(z_j) + kn_{av,\beta} z_j - \pi m z_j/d)}} \nonumber \\
 \label{eq:aa02}
\end{eqnarray}
and $n_e(z) \equiv \sqrt{ n^2(z) -\beta^2}$. Therefore 
 \begin{eqnarray}
v_1(z) &=&-{i\over {4\pi}} \sum_j \ e^{\displaystyle{2i(-\psi(z_j) 
+ k n_{av,\beta} z_j- \pi q z_j)/d}} \nonumber \\
& \ln&{{n_e(z_j+0)}\over{n_e(z_j-0)}} \ \sum_{m \ne q} 
{{ e^{\displaystyle{ i 2 \pi (m-q)(z-z_j)/d}} }\over {m-q-\delta_q d/\pi}} \ . 
 \label{eq:aa03}
\end{eqnarray}
Using eqs. 1.445.7 and 8 of \cite{Grad},  we find
 \begin{eqnarray}
v_1(z) &=& -{i\over {4}} \sum_j \ln{{n_e(z_j+0)}\over{n_e(z_j-
0)}} \ e^{{2i(-\psi(z_j)+ k n_{av,\beta} z_j- \pi q z_j/d)}} \nonumber \\ 
&.&   \left[  {1\over{\delta_q d}} -{{e^{{i[(2({\bar z}-z_j)/d-
1)\delta_q d]}} }\over {\sin \delta_q d}} \right]  \ , 
 \label{eq:aa04}
\end{eqnarray}
with ${\bar z} = z$ when $0 < z-z_j < d$ and ${\bar z} = z + d$ when $z-z_j
<0$. 
\\ 
For the asymmetric cell of the photonic crystal of interest and 
$z=0+$, this gives: 
 \begin{eqnarray}
u = v_1(0+) &=& -{i\over 4} \ln{{n_{2\beta}}\over{n_{1\beta}}} 
\bigg[ e^{i\phi_1} \left( {1\over {\delta_q d}} -{{e^{i (1-
2d_t/d)\delta_q d}}\over{\sin(\delta_q  d)}} \right) \nonumber \\  
&-&e^{i\phi_2}\left( {1\over {\delta_q d}} - {{e^{i (1-
2(d_2+d_t)/d)\delta_q d}} \over {\sin(\delta_q d)}} \right)\bigg]~. 
\label{eq:aa05}
\end{eqnarray}
with $\phi_1$ and $\phi_2$ given in eq. \ref{eq:st47}.

\subsection{ Analytic expression for $c_2$ }

According to eq. 18 of \cite{G1}: 
 \begin{equation}
c_2 \equiv {{id}\over {2\pi}} \sum_{m\ne q} {{|s_m|^2}\over {m-q- 
\delta_q d/\pi}} \ , 
 \label{eq:aa06}
\end{equation}
and using eqs. \ref{eq:st46} and \ref{eq:st47}, we have 
 \begin{eqnarray}
c_2 &=&  {i\over {2\pi d}} \left(\ln\left({n_{2,\beta}\over 
n_{1,\beta}} \right)\right)^2 \ S_2  \qquad {\rm with} \nonumber \\   
S_2 &\equiv& \sum_{m \ne q} {{ \sin^2(Pm+T) }\over{m-q-\epsilon}} \ , 
 \label{eq:aa07}
\end{eqnarray}
where
 \begin{equation} 
P \equiv \pi{d_2\over d} \quad , \quad T \equiv {{k d_1 d_2}\over d} 
(n_{2,\beta} - n_{1,\beta}) 
 \label{eq:aa08}
\end{equation}
and $\epsilon = \delta_q d/ \pi$.
Using eqs. 1.445.6 to 8 of \cite{Grad} we arrive at:
 \begin{eqnarray}
 S_2 &=& {{\sin^2 C}\over \epsilon} + \pi {{\sin(P\epsilon +C) \ 
\sin((\pi-P)\epsilon -C)}\over {\sin \pi \epsilon}} \ , \nonumber \\  
 \label{eq:aa09}
\end{eqnarray}
with $C\equiv P q +T$. Inserting this result into eq. \ref{eq:aa07} 
we have the desired analytic expression for $c_2$.

\begin{onecolumn}

\begin{figure}
\leavevmode
\epsfxsize=13cm
\epsffile{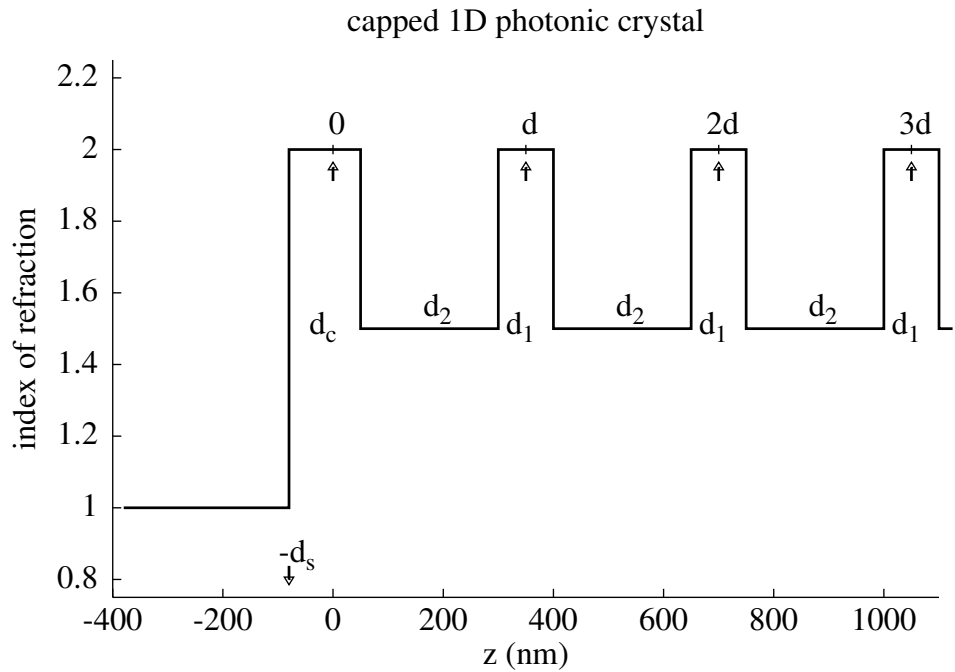}
\caption{Sketch of a typical 1D PC, for the case used in 
calculations ($n_1 = 2.0$, $d_1 = 100$ nm, $n_2 = 1.5$, $d_2 = 250$ nm,) 
A symmetric unit cell was chosen, and $d_c = 80$ nm.} 
 \label{fig01}
\end{figure}

\begin{figure}
\leavevmode
\epsfxsize=13cm
\epsffile{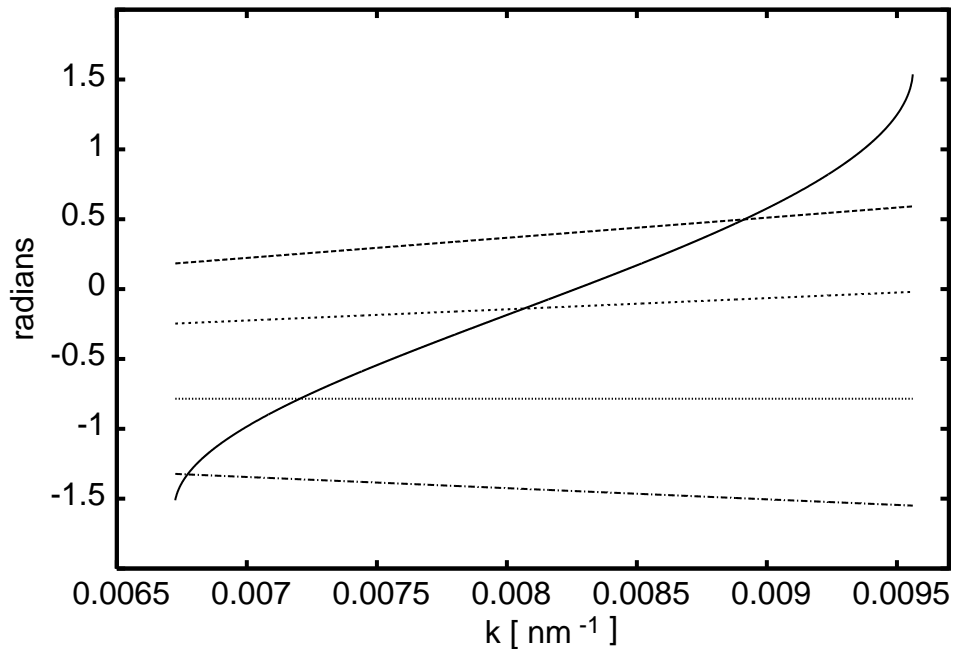}
\caption{Graphical solution of eq. \protect{\ref{eq:st23}} for $n_0 = 1$, $n_1 
= 2$, $d_1 = 100 $ nm, $n_2 = 1.5$, $ d_2 = 250 $ nm.  and $\beta = 
1.2$.  Continuous line: $\theta_{\lambda -A}$; 
dashed straight lines: r.h.s. for several values, from top to bottom, 
$d_c = 0.05, 0.25, 0.50$ and $0.75 d_1$.   } 
 \label{fig02}
\end{figure}

\begin{figure}
\leavevmode
\epsfxsize=13cm
\epsffile{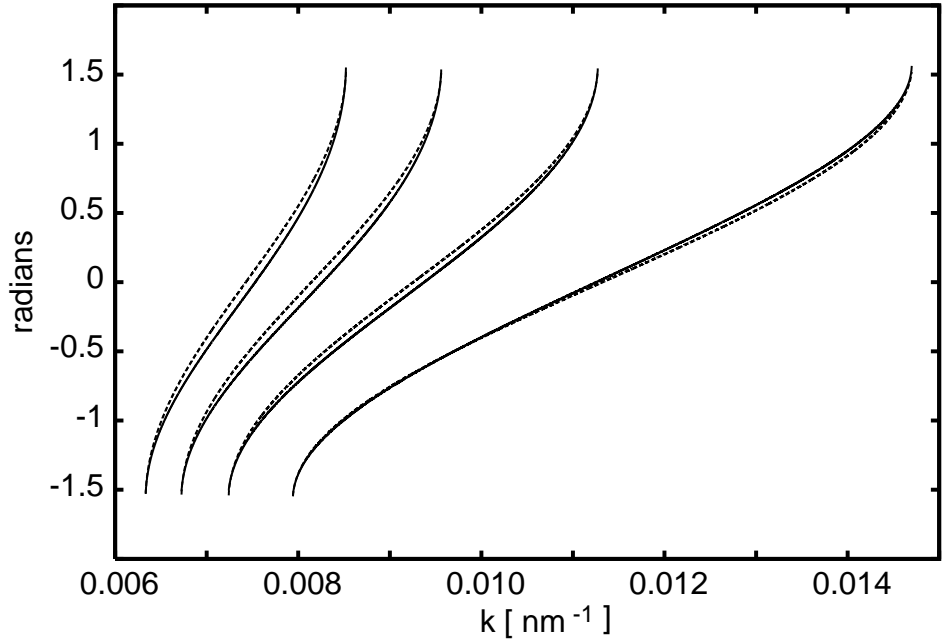}
\caption{Exact (continuous line) v.s. empirical approximation 
(dashed line) to the argument of $\lambda -A$. From left to right, 
$\beta = 1.1, 1.2, 1.3 $ and $1.4$ } 
 \label{fig03}
\end{figure}

\begin{figure}
\leavevmode
\epsfxsize=13cm
\epsffile{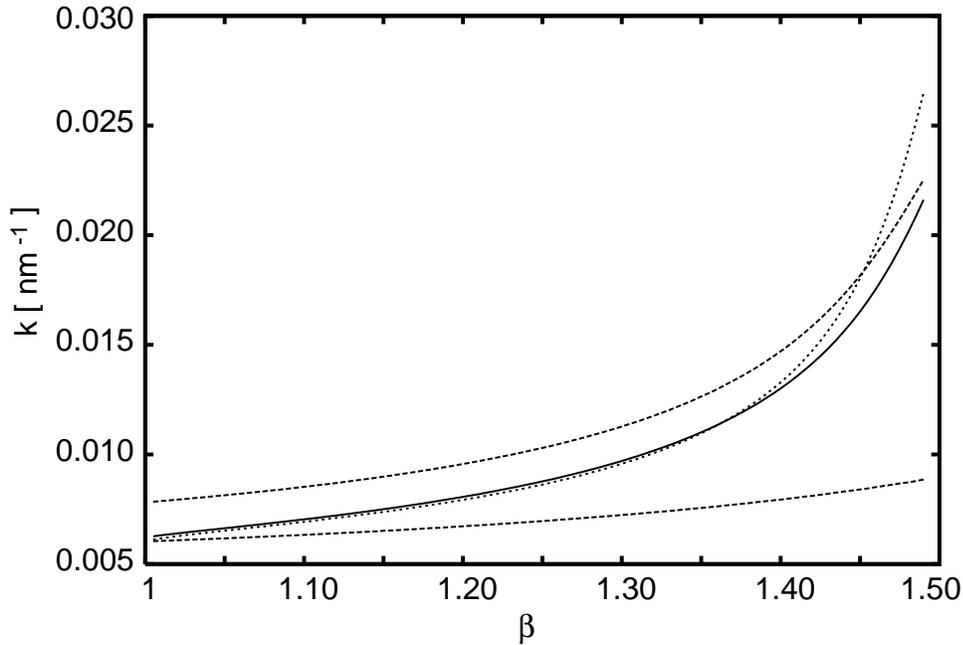}
\caption{Dispersion relation $k = k(\beta)$ when $d_c = 0.25 d_1$. Continuous
line: exact solution of eq. \protect{\ref{eq:st16}}. Dotted line: linear
approximation, eq. \protect{\ref{eq:st26}}, based on the 
empirical form of arg($\lambda -A$).  
The dashed lines are the boundaries of the first bandgap. } 
 \label{fig04}
\end{figure} 

\begin{figure}
\leavevmode
\epsfxsize=13cm
\epsffile{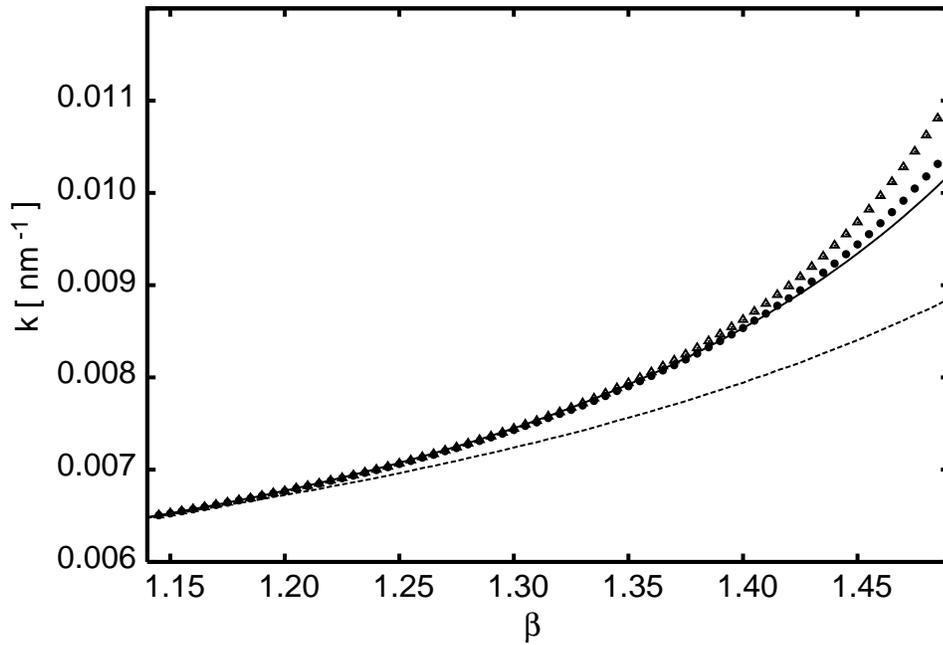}
\caption{Dispersion relation when $d_c= 0.75 d_1$ . Exact 
(continuous line) v.s. approximations based on the empirical form of 
arg($\lambda -A$). Triangles: eq. \protect{\ref{eq:st32}}; 
filled circles: eq. \protect{\ref{eq:st31}}. The dashed line is the 
lower boundary of the first bandgap. } 
 \label{fig05}
\end{figure}

\begin{figure}
\leavevmode
\epsfxsize=13cm
\epsffile{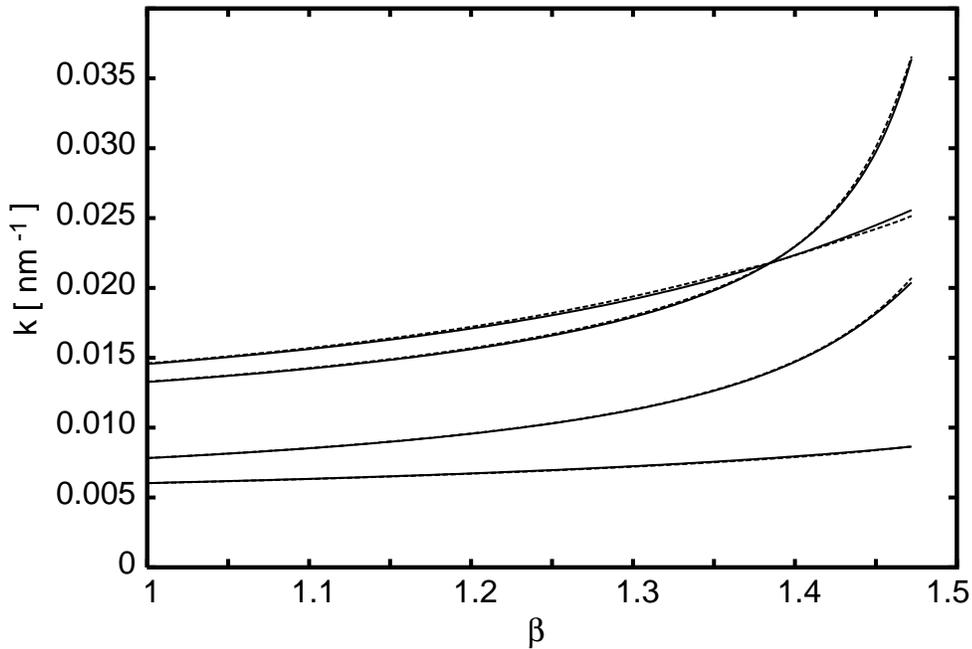}
\caption{First and second bandgap boundaries: Thin continuous 
lines: exact. Dashed lines: second semiclassical approximation 
described in text. } 
 \label{fig06}
\end{figure}

\begin{figure}
\leavevmode
\epsfxsize=13cm
\epsffile{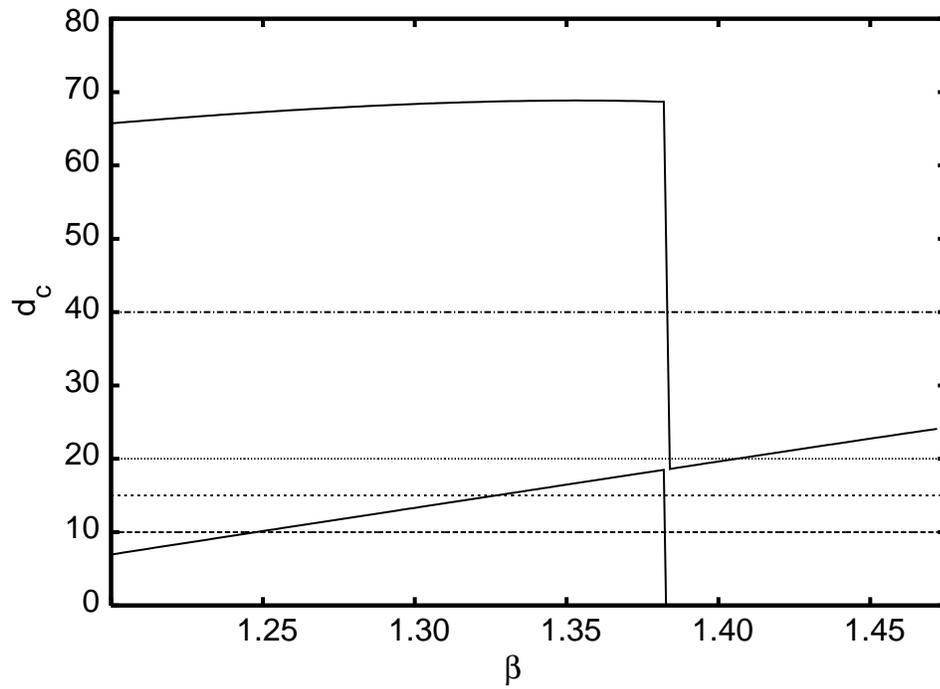}
\caption{Second bandgap: Continuous lines $d_{c,min}$ and 
$d_{c,max}$ as predicted from eq. \protect{\ref{eq:st33}}. 
Dashed horizontal lines: from bottom to top $d_c = 10, 15, 20$ and $ 
40$  nm. } 
 \label{fig07}
\end{figure}

\begin{figure}
\leavevmode
\epsfxsize=13cm
\epsffile{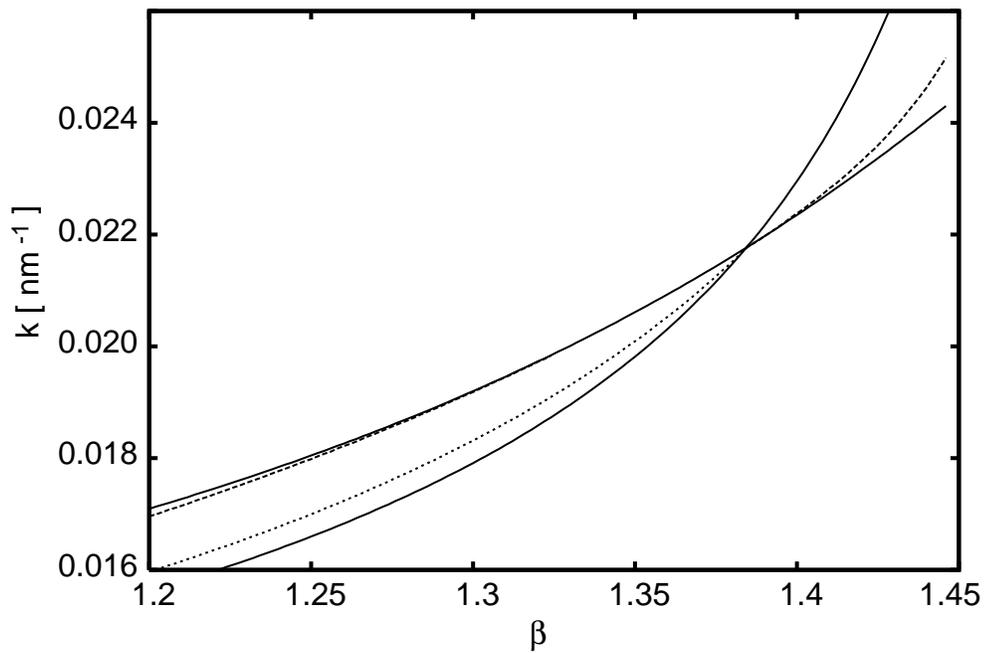}
\caption{Second bandgap: Continuous lines: zone boundaries. 
Dispersion relation for $d_c = 0.15 d_1$ (dashed line), and  $d_c 
= 0.40 d_1$ (dotted). } 
 \label{fig08}
\end{figure}

\begin{figure}
\leavevmode
\epsfxsize=13cm
\epsffile{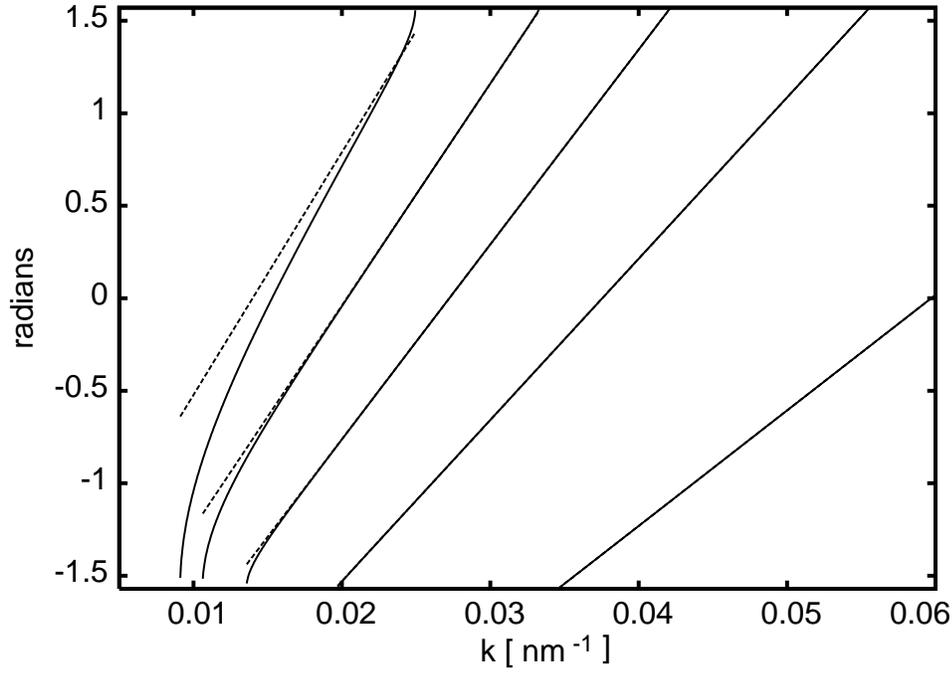}
\caption{Argument of $\lambda - A$ in first bandgap.  From left to 
right: $\beta = 1.51, 1.6, 1.7, 1.8$ and $1.9$. Continuous lines: 
exact; dashed lines: linear approximation of eq. \protect{\ref{eq:st42}} 
} 
 \label{fig09}
\end{figure} 

\begin{figure}
\leavevmode
\epsfxsize=13cm
\epsffile{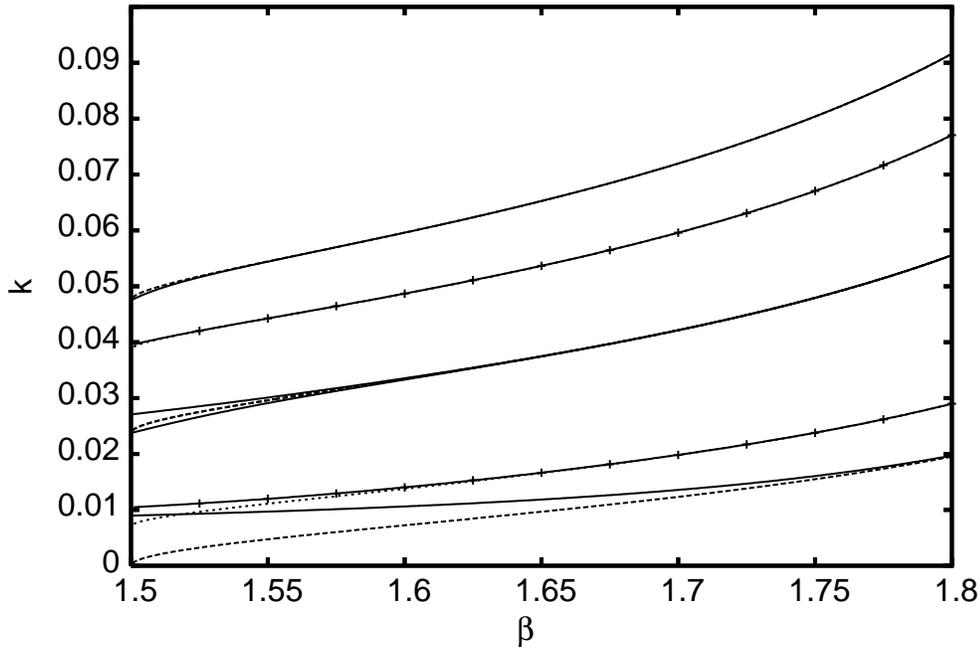}
\caption{First and second bandgaps when $\beta > n_2$. 
Continuous lines: exact boundaries. Dashed lines: approximation described 
in text. Surface waves: solid line with $+$ symbols: 
exact; dotted lines: approximation given in the text. } 
 \label{fig10}
\end{figure}

\begin{figure}
\leavevmode
\epsfxsize=13cm
\epsffile{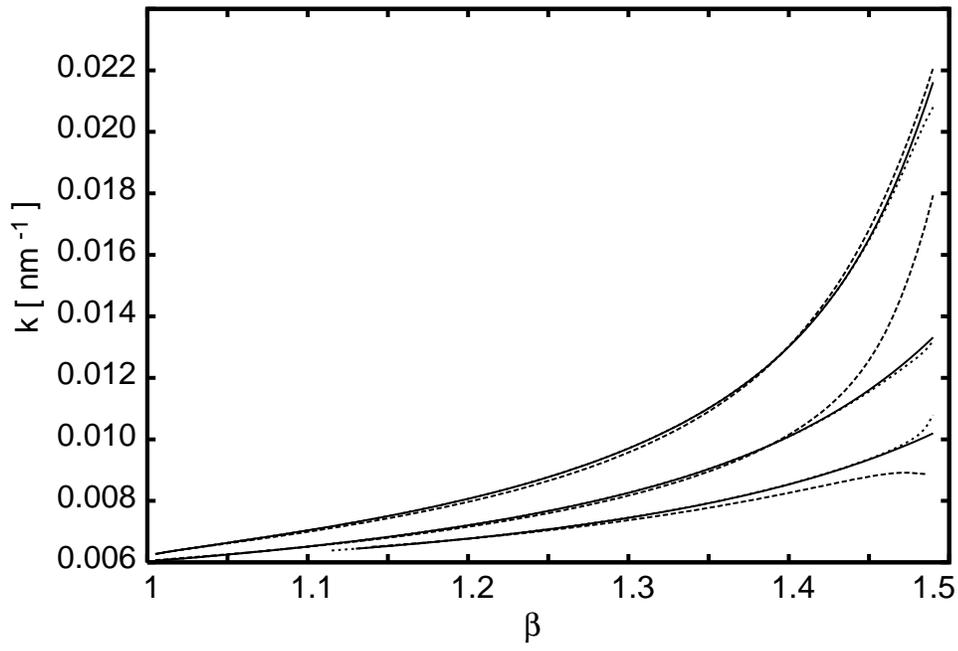}
\caption{First bandgap: $ k = k(\beta)$ curves for $d_c = 25, 50 $ and 
$75$ nm. Continuous lines: exact. Long dashes: first 
approximation, short dashes: second approximation. The latter curves 
are so close to the exact ones that the difference can be seen only 
when $\beta > 1.45$. } 
 \label{fig11}
\end{figure}

\end{onecolumn}

\end{document}